\documentclass{article}

\usepackage{amssymb,amsfonts,amsmath,stmaryrd}
\usepackage{cite,enumerate,float,indentfirst}
\usepackage{color}
\usepackage{tikz}
\usetikzlibrary{arrows,snakes,backgrounds}

\def\be{\begin{eqnarray}}
\def\ee{\end{eqnarray}}
\def\nn{\nonumber}

\def\p{\partial}
\def\tr{{\rm tr}\,}
\def\Tr{{\rm Tr}\,}

\definecolor{red}{rgb}{1,0,0}
\definecolor{orange}{rgb}{1,0.5,0}
\definecolor{violet}{rgb}{0.7,0,1}

\hoffset= - 1.0in         

\begin{document}

\begin{center}
\begin{small}
\hfill FIAN/TD-09/21\\
\hfill IITP/TH-13/21\\
\hfill ITEP/TH-16/21\\
\hfill MIPT/TH-12/21\\
\end{small}
\end{center}

\vspace{.5cm}

\begin{center}
\begin{Large}\fontfamily{cmss}
\fontsize{17pt}{27pt}
\selectfont
	\textbf{Non-Abelian W-representation for GKM}
	\end{Large}
	
\bigskip \bigskip

\begin{large}A. Mironov$^{a,b,c,}$\footnote{mironov@lpi.ru; mironov@itep.ru},
V. Mishnyakov$^{d,a,b,}$\footnote{mishnyakovvv@gmial.com},
A. Morozov$^{d,b,c,}$\footnote{morozov@itep.ru}
 \end{large}
\\
\bigskip

\begin{small}
$^a$ {\it Lebedev Physics Institute, Moscow 119991, Russia}\\
$^b$ {\it ITEP, Moscow 117218, Russia}\\
$^c$ {\it Institute for Information Transmission Problems, Moscow 127994, Russia}\\
$^d$ {\it MIPT, Dolgoprudny, 141701, Russia}
\end{small}
 \end{center}

\bigskip

\begin{abstract}
$W$-representation is a miraculous possibility to define a
non-perturbative (exact) partition function as an exponential action of somehow integrated Ward identities on unity.
It is well known for numerous eigenvalue matrix models,
when the relevant operators are of a kind of
$W$-operators: for the Hermitian matrix model with the Virasoro constraints,
it is a $W_3$-like operator, and so on.
We extend this statement to the monomial generalized Kontsevich models (GKM),
where the new feature is appearance of an ordered P-exponential
for the set of non-commuting operators of different gradings.
\end{abstract}

\bigskip

\section{Introduction. Hermitian model and the idea of $W$-representation}

Partition function of matrix models \cite{UFN3} usually satisfies an
exhaustive set of Virasoro and $W$-constraints,
which are, however, not so easy to solve.
For example, for the Hermitian matrix model with the partition function $Z_N\{p\}$ where $N$ is the size of matrix, the Ward identities read \cite{Vircon}
\be
\hat L_nZ_N\{p\} = 0,
\ \ \ \ n\geq -1
\ee
and the operators
\be
\hat L_n:=\sum_{k} (k+n)p_k\frac{\p}{\p p_{k+n}} + \sum_{a=1}^{n-1} a(n-a)\frac{\p^2}{\p p_a\p p_{n-a}}
+ 2Nn\frac{\p}{\p p_n}    + N^2\delta_{n,0} + Np_1 \delta_{n+1,0} - \underline{(n+2)\frac{\p}{\p p_{n+2}}}
\label{virc}
\ee
form a Borel subalgebra of the Virasoro algebra. The underlined term breaks the grading, the grading of $p_k$ being $k$. Such a choice of this term corresponds to choice of the Gaussian phase. In this phase, this system of equations has a unique solution \cite{AMM,Max1},
which is given by \cite{MSh} (see \cite{wrep,wrep1,wrep2} for early precursors)
\be
Z\{p\} = e^{\hat O_{2}} \cdot 1
\label{WHerm}
\ee
where
\be
\hat O_{2} := {1\over 2} \sum_{k,n} (k+n)p_{n+2}p_k\frac{\p}{\p p_{k+n}} +{1\over 2} \sum_{a,b} ab p_{a+b+2}\frac{\p^2}{\p p_a\p p_b}+N\sum_{k=3} (k-2)p_k \dfrac{\partial }{\partial p_{k-2}}
+ {N p_1^2\over 2}+{N^2 p_2\over 2}
\ee
As explained in \cite{MMMR2} (see also \cite{Max2}), this representation can be deduced from the fact that the Virasoro constraints
(\ref{virc}) can be all encoded in a single equation
\be
\sum_{n\geq 1} p_n\hat L_{n-2}Z_N\{p\}=\left(2\hat O_{2}
- \overbrace{\underline{\sum_{n\geq 1} np_n\frac{\p}{\p p_{n}}}}^{\hat l_0} \right) Z\{p\} = 0
\label{eqnHerm}
\ee
that has a unique solution. The operators commute in the simple way:
\be
\left[\hat l_0, \hat O_{2}\right] = 2 \hat O_{2}
\ee
and
\be
\hat l_0 \cdot 1 = 0
\ee
i.e. $\hat l_0$ is the grading operator, and the grading of $\hat O_{2}$ is 2.
This is the main point: we combined Virasoro constraints in such a way
that the grading-breaking piece is converted into grading operator $\hat l_0$.
Now introducing the grading parameter $x$ via the rescaling $p_k\to x^kp_k$, one comes to the equation
\be
\left(-x\frac{d}{dx} + 2x^2\hat O_2\right) Z = 0
\ee
with an obvious solution
\be\label{O2}
Z \sim \exp\left(\hat O_2x^2\right)
\ee
Since the solution is unique, one establishes that (\ref{O2}) provides a representation of the Hermitian matrix model partition function.
With our operators we do not need $x$, and (\ref{eqnHerm}) just has (\ref{WHerm}) as
an obvious solution.

Since \cite{MSh}, there were many more examples of $W$-representations for many different models
\cite{otherW}, see \cite{MMMR2} for a recent summary and for an evidence for unambiguity of
solutions.
However, there remains an important exception from the general list:
the monomial Kontsevich models \cite{GKM,versus} beyond the simplest cubic example \cite{Alexcubic}.
The goal of this letter is to fill the gap and provide a simple description of
what happens to $W$-representation for the generalized Kontsevich model (GKM).

The partition function of the monomial GKM is given by the matrix integral over $N\times N$ Hermitian matrix $X$,
\be\label{GKM}
Z_r(M):={\cal N}_r\cdot\int dX e^{-{\Tr X^{r+1}\over r+1}+\Tr M^rX}\\
{\cal N}_r={e^{-{r\over r+1}\Tr M^{r+1}}\over \int dXe^{-{1\over r}\sum_{a+b=r-1}\Tr M^aXM^bX}}\nn
\ee
and depends on the external matrix $M$. At large $M$ (so called Kontsevich phase \cite{GKMU}),
this partition function is understood as a power series in time-variables $p_k:=\Tr M^{-k}$, the coefficients of this power series being independent of the size of matrix $N$. Hence, the notation $Z_r(M)=Z_r\{p\}$. This partition function does not depend on $p_{rk}$-variables, and is normalized so that $\lim_{M\to\infty}Z_r(M)=Z_r\{0\}=1$.

The Ward identities of this matrix model are described by constraints from the $W^{(r)}$-algebra and become rather involved at large $r$. In the next section, we consider the simplest case of $r=2$, when they form a Borel subalgebra of the Virasoro algebra. In section 4, we consider the first non-trivial case of $r=3$, when the $W$-algebra constraints emerge for the first time, and, in section 5, we consider the generic $r$ case.

\section{W-representation in cubic case}

We start with the partition function (\ref{GKM}) with $r=2$ \cite{Kon}. In this case, the partition function satisfies the Virasoro constraints  \cite{MMM,W,AMMP},
\be
\phantom{.}_{2}\hat{ L}_n  Z_{2}\{p\} = 0, \ \ \ \ n\geq -1
\label{virK}
\ee
\be
\phantom{.}_{2}\hat { L}_n :={1\over 2}\sum_{k\in{\tiny odd}} (k+2n)p_k\frac{\p}{\p p_{k+2n}} + {1\over 4}\sum_{a,b\in{\tiny odd}}^{a+b=2n} ab\frac{\p^2}{\p p_a\p p_b}
+ {p_1^2\over 4}\delta_{n,-1}+{1\over 16}\delta_{n,0}-\underline{{(2n+3){\p\over\p p_{2n+3}}}}
\label{VirK}
\ee
Here the sums over $k$ and $a$ run over odd numbers. These constraints can be encoded in a single equation that has a unique solution,
\be\label{EqK}
\sum_{n=1}p_{2n-1}\cdot\phantom{.}_{2}\hat L_{n-2}Z_{2}\{p\}=0
\ee
This equation contains the terms of gradings 0 and 3. The zero grading term comes from the last
(underlined) term in (\ref{VirK}) and is
\be
\hat l_0 = \sum_{k\in{\tiny odd}} kp_k\frac{\p}{\p p_k}
\ee
so that (\ref{EqK}) takes the form
\be
\left(\hat l_0-3\hat O_3\right)Z_{2}\{p\}=0
\ee
with the operators of grading 3 being
\be
\hat O_3:=\dfrac{1}{6} \sum_{k,l\in{\tiny odd}}(k+l-3) p_kp_l\dfrac{\partial}{\partial p_{k+l-3}}
+\dfrac{1}{12}\sum_{k,l\in{\tiny odd}} (k-l-3)l p_{k}\dfrac{\partial^2}{\partial p_{k-l-3}\partial p_l}
+\dfrac{1}{48}p_3+\dfrac{1}{12}p_1^3
\ee
The commutation relation is
\be
\left[\hat l_0, \hat O_3\right] = 3\hat O_3
\ee
Introducing the grading parameter $x$ via the rescaling $p_k\to x^kp_k$, we come to the equation
\be\label{eq3}
\left(-x\frac{d}{dx} +3x^3\hat O_3\right) Z_{2}\{p\} = 0
\ee
Its solution is exponential,
\be
Z_{2}\{p\}=\exp\left(x^3\hat O_3\right)\cdot 1
\ee
which is nothing but the standard $W$-representation \cite{Alexcubic,MMMR2}.

\section{W-representation in quartic case}

Now we consider the $r=3$ case. This is the first truly non-trivial case.
We have now a combination of Virasoro and $W$-constraints \cite{FKN1,Mikh,MP,KMMMP}
\be
\phantom{.}_{3}\hat{ L}_n  Z_{3}\{p\} = 0, \ \ \ \ n\geq -1\nn\\
\nn\\
\phantom{.}_{3}\hat{ W}_n^{(3)}  Z_{3}\{p\} = 0, \ \ \ \ n\geq -2
\label{virGKM}
\ee
\be
\phantom{.}_{3}\hat { L}_n :={1\over 3}\sum_{k} (k+3n)p_k\frac{\p}{\p p_{k+3n}} + {1\over 6}\sum_{{a,b=1}\atop{a+b=3n}} ab\frac{\p^2}{\p p_a\p p_{b}}
+ {p_1p_2\over 3}\delta_{n,-1}+{1\over 9}\delta_{n,0}-\underline{(3n+4){\p\over\p p_{3n+4}}}
\nn\\
\phantom{.}_{3}\hat { W}_n ^{(3)}:={1\over 9}\sum_{k,l=1}(k+l+3n)P_kP_l{\p\over\p p_{k+l+3n}}+{1\over 9}\sum_{k=1}\sum_{{a,b=1}\atop{a+b=k+3n}}abP_k{\p^2\over
\p p_a\p p_{b}}+{1\over 27}\sum_{{a,b,c=1}\atop{a+b+c=3n}}abc{\p^3\over\p p_a\p p_b\p p_{c}}+\nn\\
+{1\over 27}\sum_{{a,b,c=1}\atop{a+b+c=-3n}}P_aP_bP_{c}
\label{VirGKM}
\ee
where $P_k:=p_k-3\cdot\delta_{k,4}$, and $a,b,c,k,l$ in the sums are not divisible by 3.
They can be combined into a single equation that unambiguously determines the partition function
\be\label{seq4}
\sum_{n=1}p_{3n-1}\cdot\phantom{.}_{3}\hat { W}_{n-3}^{(3)}Z_{3}\{p\}+c
\sum_{n=1}p_{3n-2}\cdot\phantom{.}_{3}\hat { L}_{n-2}Z_{3}\{p\}=0
\ee
where the parameter $c$ can be chosen rather arbitrarily (only non-negative rational $c$ can give rise to additional superfluous solutions of this equation) \cite{MMMR2}. At the l.h.s. of this equation, there are operators of
gradings 0, 4 and 8.

For the special choice of  $c=-1$,
the coefficients in front of the sum $\sum_n(3n-1)p_{3n-1}\frac{\p}{\p p_{3n-1}}$,
coming from the first term in (\ref{seq4}),
and in front of the sum $\sum_n(3n-2)p_{3n-2}\frac{\p}{\p p_{3n-2}}$, coming from the second term,
are equal to each other,
so that the zero grading operator is nothing but $\hat l_0$
\be
\hat l_0 = \sum_k kp_k\frac{\p}{\p p_k}
\ee
with $k$ not divisible by 3. With this choice, (\ref{seq4}) looks like
\be
\left(\hat l_0-4 \hat O_4- 8\hat O_8\right)Z_{3}\{p\}=0
\ee
where the operators of gradings 4 and 8 are
\be
\hat O_4:={1\over 12}\sum_{n=1} p_{3n-1}\left(
2\sum_{k }(k+3n-5) p_k{\p\over\p p_{k+3n-5}}+ \sum_{a+b=3n-5}ab
{\p^2\over \p p_a\p p_b}\right)+\nn\\
+{1\over 24}\sum_{n=1} p_{3n-2}\left(
2\sum_{k } (k+3n-6) p_k{\p\over\p p_{k+3n-6}}
+ \sum_{a+b=3n-6}ab {\p^2\over\p p_a\p p_b}\right)
+{p_4\over 36}+{p_1^2p_2\over 6}
\ee
and
\be
\hat O_8:=-{1\over 8\cdot 27}\left\{\sum_{n=1}p_{3n-1}\left(
3\sum_{k,l}(k+l+3n-9)p_kp_l{\p\over\p p_{k+ l+3n-9}}
+3\sum_{k=1}
\sum_{{a,b=1}\atop{a+b=k+3n-9}}ab p_k{\p^2\over
\p p_a\p p_{b}}
\right.\right. +\nn\\
+\left.\left.\sum_{{a,b,c=1}\atop{a+b+c=3n-9}}abc{\p^3\over\p p_a\p p_b\p p_{c}}\right)
+p_1^3p_5+p_2^4+3p_1^2p_2p_4\right\}
\ee
and the sums over $k$, $l$, $a$, $b$, $c$ run over positive integers not divisible by 3.

\bigskip

The commutation relations are
\be
\left[\hat l_0, \hat O_4\right] = 4\hat O_4, \ \ \ \
\left[\hat l_0, \hat O_8\right] = 8\hat O_8
\ee
Introducing the grading parameter $x$ via the rescaling $p_k\to x^kp_k$, we come to the equation\footnote{
In the case of arbitrary $c$ in (\ref{seq4}),
one has to consider two different gradings, $p_{3k-1}\to x^{3k-1}p_{3k-1}$ and $p_{3k-2}\to (x\alpha)^{3k-2}p_{3k-2}$ which leads to the equation
$$
\left(-x\frac{d}{dx} +(1+c)\alpha\frac{d}{d\alpha} +4x^4\hat O_4(\alpha,c)+8x^8\hat O_8(\alpha)\right) Z_{3}\{p\} = 0
$$
with the operators $\hat O_{4,8}$ depending on $\alpha$ and $\hat O_{4}$, on the constant $c$. Another possibility is to define yet another operator of zero grading, $\hat O_0:=\sum_k(3k-2)p_{3k-2}{\p\over\p p_{3k-2}}$ and deal with the equation
$$
\left(-x\frac{d}{dx} +(1+c)\hat O_0 +4x^4\hat O_4(c)+8x^8\hat O_8\right) Z_{3}\{p\} = 0
$$
It makes the whole consideration more involved. In particular, at some peculiar rational values of $c$, there is a degeneration, which gives rise to additional superfluous solutions to Eq.(\ref{seq4}).  For instance, at $c=+2$, one gets
\be
Z_{3}\{p\}=1+\alpha\cdot p_1p_2+{1\over 6}p_1^2p_2+{1\over 36}p_4+\ldots
\ee
and the coefficient $\alpha$ is not determined from Eq.(\ref{seq4}).
}
\be\label{eq4}
\left(-x\frac{d}{dx} +4x^4\hat O_4+8x^8\hat O_8\right) Z_{3}\{p\} = 0
\ee
Its solution is going to be an {\it ordered} exponential
\be\label{Wrep}
Z_{3}\{p\} = P\exp\left(\int^x \Big(4x'^4\hat O_4 +8x'^8 \hat O_8\Big)\frac{dx'}{x'}\right)\cdot 1
=\nn \\
= 1 + \int^x \Big(4x'^4\hat O_4 +8x'^8 \hat O_8\Big)\frac{dx'}{x'} \cdot 1
+ \int^{x} \Big(4x'^4\hat O_4 +8x'^8 \hat O_8\Big)\frac{dx'}{x'} \int^{x'} \Big(4x''^4\hat O_4 +8x''^8 \hat O_8\Big)\frac{dx''}{x''} \cdot 1 + \ldots
= \nn \\
= \left(1 + x^4\hat O_4 + x^8\left({1\over 2}\hat O_4^2+\hat O_8\right)+x^{12}\left({1\over 6}\hat O_4^3
+{1\over 3}\hat O_4\hat O_8+{2\over 3}\hat O_8\hat O_4\right)+\ldots\right)\cdot 1
\ee
The simplest way to generate this expansion is as follows. Let us look for a solution in the form $Z_{4}\{p\}=\sum_k x^{4k}\hat\Psi_k\cdot 1$. Then, (\ref{eq4}) is equivalent to the recursion relation
\be
\hat\Psi_k={1\over k}\hat O_4\hat \Psi_{k-1}+{2\over k}\hat O_8\hat \Psi_{k-2}
\ee
with the initial conditions $\hat\Psi_0=1$, $\hat\Psi_1=\hat O_4$.

\bigskip

Note that the recursion relation is consistent with similar relations obtained by J.Zhou \cite{Zhou}, though we derive them within a different framework. However, the operators $\hat O_4$ and $\hat O_8$ do not commute and, hence, do not lead to a simple exponential $W$-representation form of (\ref{Wrep}) (this is not quite consistent with the conclusion of \cite{Zhou}):
\be
P\exp\left(x^4\hat O_4 +x^8 \hat O_8\right)=\exp\left(x^4\hat O_4 +x^8 \hat O_8-{x^{12}\over 6}[\hat O_4,\hat O_8]
-{x^{20}\over 60}[\hat O_8,[\hat O_4,\hat O_8]]-{x^{20}\over 360}[\hat O_4,[\hat O_4,[\hat O_4,\hat O_8]]]
+\ldots\right)
\ee

The series (\ref{Wrep}) is one of the most effective technical ways to generate the partition function $Z_{3}\{p\}$ as an expansion in  powers of $p_k$'s (see the associated data and Appendix B, where, as an illustration, we evaluate $Z_{3}\{p\}$ up to $x^{16}$).

\section{W-representation for arbitrary monomial potential}

\subsection{The case of $r=4$ }

Let us briefly sketch the next $r=4$ case.
This time we should use the following constraints:
\begin{equation}
\begin{split}
\phantom{.}_{4}\hat{ L}_n Z_4\{p\} = 0&, n \geq -1
\\
\phantom{.}_{4}\hat{W}_n^{(3)} Z_4 \{p\} = 0&, n \geq -2
\\
\phantom{.}_{4}\hat{W}_n^{(4)} Z_4\{p\} = 0&, n \geq -3
\end{split}
\end{equation}
The corresponding $W$ algebra can be expressed in terms of bosonic currents:
\be\label{W2}
\phantom{.}_{4}\hat{ L}_n= \dfrac{1}{8}\sum_{n_1+n_2 = 4n}  : J_{n_1} J_{n_2}:+
\ \dfrac{5}{32}\delta_{n,0}
\ee
\be\label{W3}
\phantom{.}_{4}\hat{W}_n^{(3)}= \dfrac{1}{48}\sum_{n_1+n_2+n_3 = 4n}  : J_{n_1} J_{n_2}J_{n_3} :
\ee
\be\label{W41}
\phantom{.}_{4}\hat{W}_n^{(4)} = \dfrac{1}{256}\sum\limits_{n_1+n_2+n_3+n_4 = 4n}  : J_{n_1} J_{n_2}J_{n_3}J_{n_4} :  - \dfrac{1}{128}  \sum\limits_{ \substack{
p+q=n
\\n_1+n_2 =4p
\\n_3+n_4=4q} } : J_{n_1} J_{n_2} J_{n_3} J_{n_4}  : +\nn\\
+\dfrac{5}{512}\sum\limits_{n_1+n_2=4n}: J_{n_1} J_{n_2}:-\dfrac{1}{256}\sum\limits_{n_1+n_2=4n}
(n_1)_r(n_2)_r\, :J_{n_1} J_{n_2}:
- \ \dfrac{9}{4096}\delta_{n,0}
\ee
where $(n)_r$ denotes $n$ modulo $r$.
The currents are:
\be\label{sh5}
\left\{
\begin{array}{rcl}
J_{-n}&=&p_{n} - 4 \delta_{n,5}\cr
&&\hspace{3cm}n>0\cr
J_n&=&n \dfrac{\partial}{\partial p_n}
\end{array}
\right.
\ee
and the normal ordering implies all the derivatives moved to the right.
The sums in these expressions run over integers not divisible by 4.
The last term in $\phantom{.}_{4}\hat{W}_n^{(4)}$ comes from the anomaly.
Notice
a misprint in $\phantom{.}_{4}\hat{W}_n^{(4)}$ of \cite[Appendix C]{FKN2}.

\bigskip

As usual \cite{MMMR2} we consider a peculiar linear combination of these constraints:
\begin{equation}
c_3 \sum_{n=1} p_{4n-1} \cdot \phantom{.}_{4}\hat{W}_{n-4}^{(4)} Z_4\{p\} + c_2 \sum_{n=1} p_{4n-2}\cdot \phantom{.}_{4}\hat{W}_{n-3}^{(3)} Z_4\{p\} +c_1 \sum_{n=1} p_{4n-3} \cdot \phantom{.}_{4}\hat{ L}_{n-2} Z_4\{p\} =0
\end{equation}
and according to \cite{MMMR2} this equation has a unique solution for almost arbitrary constants $c_i$.
For our current purposes they can be chosen so
that the zero grading operators combine into $l_0 = \sum p_k \dfrac{\partial}{\partial p_k}$.
This choice is
\begin{equation}\label{ci}
c_i = (-1)^i
\end{equation}
It deserves making a brief remark on grading.
If we neglect the shift of the fifth time in (\ref{sh5}), then all the terms coming from $\phantom{.}_{4}\hat{W}_{n-4}^{(4)}$ have the grading 15.
The third term in (\ref{W41}) contains at maximum one shift, which means
there is also a term of grading 10.
The second term contains terms with one or two shifts, which means there are terms of grading 5 and 10.
Hence, the zero grading terms come only from the leading term in (\ref{W41}) and, similarly, in (\ref{W2}) and (\ref{W3}). This immediately gives (\ref{ci}).

\bigskip

For this choice (\ref{ci}) the equation for the partition function acquires the form
\begin{equation}
\left(l_0 - 5\, \hat{O}_5 - 10\, \hat{O}_{10} - 15\, \hat{O}_{15}  \right) Z_{4} \{p\}=0
\end{equation}
Then the $W$-representation is given by:
\begin{equation}
Z_{4}\{p\} =P \exp \left(x^5 \hat{O}_5+ x^{10} \hat{O}_{10} + x^{15} \hat{O}_{15}\right) \cdot 1
\end{equation}
We illustrate this representation by evaluation of the partition function $Z_{4}\{p\}$ in Appendix B up to order $15$.
In the 5-th and 10-th order, available at \cite{MMQgen}, it coincides with the answer in that paper.

\subsection{Towards arbitrary $r$}

Partition function $Z_r$ in the GKM (\ref{GKM}) with potential $x^{r+1}$
does not depend on $p_{nr}$  and  satisfies the whole
set of $W$-constraints of the orders ranging from $2$ (Virasoro) to $r$ \cite{GKM},
\be
\phantom{.}_{r}\hat{W}^{(k)}_nZ_r\{p\}=0,\ \ \ \ \ \ \ k=2,\ldots,r,\ \ \ \ \ n\ge 1-k
\ee
and the $W$-generators are defined in \cite{FKN2} and in Appendix B, the first two being
\be\label{W2r}
\phantom{.}_{r}\hat{ L}_n= \dfrac{1}{2r}\sum_{n_1+n_2 = rn}  : J_{n_1} J_{n_2}:
+\ \dfrac{r^2-1}{24r}\delta_{n,0}
\ee
\be
\phantom{.}_{r}\hat{W}_n^{(3)}= \dfrac{1}{3r^2}\sum_{n_1+n_2+n_3 = rn}  : J_{n_1} J_{n_2}J_{n_3} :
\ee
They can be expressed through the $\mathbb{Z}_r$-twisted scalar fields \cite{FKN2}, and higher constraints
are rather involved, e.g. (see Appendix B)
\be\label{W4}
\phantom{.}_{r}\hat{W}_n^{(4)}=
\dfrac{1}{4 r^3}\sum\limits_{n_1+n_2+n_3+n_4 = rn}  \!\!\!\!\! \!\!\!\!\!: J_{n_1} J_{n_2}J_{n_3}J_{n_4} :
\ - \ \dfrac{1}{8 r^2}  \sum\limits_{ \substack{
p+q=n
\\n_1+n_2 =rp
\\n_3+n_4=rq} }  \!\!\!\!\!: J_{n_1} J_{n_2} J_{n_3} J_{n_4}  :
\!\!
+\dfrac{(r^2-1)(r-6)}{48r^2}\!\!\!\!\!\!\sum\limits_{n_1+n_2=rn}
\!\!\!\! : J_{n_1} J_{n_2}: - \nn
\ee
\be
\, -\, \dfrac{1}{8r^3}\sum\limits_{n_1+n_2=rn}
\!\!\!\!\!\! \Big(\langle n_1\rangle_r+\langle n_2\rangle_r\Big)\, :J_{n_1} J_{n_2}:
-\ \displaystyle{{(r^2-1)(r-2)(r-3)(5r+7)\over 5760r^3}\ \delta_{n,0}}
\ee
and
\be
\label{W5}
\phantom.{_r}\hat W_n^{(5)}={1\over 5r^4}
\sum\limits_{n_1+n_2+n_3+n_4+n_5 = rn}  : J_{n_1} J_{n_2}J_{n_3}J_{n_4}J_{n_5} :
- \dfrac{1}{6r^3}  \sum\limits_{ \substack{
p+q=n
\\n_1+n_2+n_3 =rp
\\n_4+n_5=rq} } : J_{n_1} J_{n_2} J_{n_3} J_{n_4} J_{n_5} :-
\ee
\be
+{(r^2-1)(r-12)\over 72r^3}\!\!\!\!\!\sum_{n_1+n_2+n_3 = rn}\!\!\!\!\!  : J_{n_1} J_{n_2}J_{n_3} :
-{1\over 6r^4}\sum_{n_1+n_2+n_3 = rn}  \Big(\langle n_1\rangle_r+\langle n_2\rangle_r+\langle n_3\rangle_r\Big): 
J_{n_1} J_{n_2}J_{n_3} :
\nn
\ee
In these formulas,
\be\label{shr}
\left\{
\begin{array}{rcl}
J_{-n}&=& p_{n} - r\mu \delta_{n,r+1}\cr
&&\hspace{3.5cm}n>0\cr
J_n&=&n \dfrac{\partial}{\partial p_n}
\end{array}
\right.
\ee
and the sums run over integers not divisible by $r$. We denote $\langle n\rangle_r:=(n)_r\Big(r-(n)_r\Big)$ where, as before, $(n)_r$ is the value of $n$ modulo $r$. At the moment, we introduce a parameter $\mu$ in the term violating the grading in order to control the grading easier. We will ultimately put $\mu=1$.

The leading term of the $W$-generator is
\be
\phantom{.}_{r}\hat { W}_n ^{(k)}:={1\over r^{k-1}} \sum_{k_i=1,k_i\notin r\mathbb{Z}}
\Big(\sum_jk_j+rn\Big)\cdot \prod_{i=1}^{r-1} \Big(p_{k_i}-r\cdot\delta_{k_i,r+1}\Big)
\cdot{\p\over\p p_{_{\sum_jk_j+rn}}}\ +\ \ldots
\ee

\bigskip

These $W$-constraints can be combined into a single equation
\be
\sum_{i=1}^{r-1} c_i \sum_n p_{nr-r+i}\cdot \phantom{.}_{r}\hat{W}^{(i+1)}_{n-i-1}\cdot Z_r\{p\} = 0
\label{eqnr1}
\ee
again with the nearly arbitrary constants $c_i$.
This equation is a sum of operators of gradings $\{(r+1)i\}$, $i=0..r-1$, which is given by the expansion of the operators
\be
\phantom{.}_{r}\hat{W}^{(i+1)}_{n}=\sum_{j=0}^i\mu^j\cdot\phantom{.}_{r}\hat{W}^{(i+1)}_{n,j}
\ee
into operators of definite gradings: $\phantom{.}_{r}\hat{W}^{(k)}_{n,j}$ has grading $rn-j(r+1)$.

\bigskip

Again the constants $c_i$ can be adjusted so that all the zero grading operators, i.e.
all the terms in
$$\sum_{n=1} p_{nr-i}\frac{\p}{\p p_{nr-i}}$$
with all $i=1,\ldots,r$,
come with the unit coefficients and combine into the grading operator
\be
\hat l_0 = \sum_{i=1}^{r-1}\sum_{n=1}^\infty  p_{nr-i}\frac{\p}{\p p_{nr-i}}
\ee
This is the choice $c_i=(-1)^i$.
Then, introducing the grading parameter $x$ via the rescaling $p_k\to x^kp_k$, we come to the equation
\be
\sum_{i=1}^{r-1} (-1)^i \sum_n p_{nr-r+i}\cdot \phantom{.}_{r}\hat{W}^{(i+1)}_{n-i-1}\cdot Z_r\{p\}
=\sum_{j=0}^{r-1}\Big(x^{r+1}\mu\Big)^j\underbrace{\sum_{i=j}^{r-1}(-1)^i\sum_{n>0,n\notin r\mathbb{Z}}
p_{nr-r+i}\cdot \phantom{.}_{r}\hat{W}^{(i+1)}_{n-i-1,j}}_{(r+1)jO_{(r+1)j}}\cdot Z_r\{p\}
=0
\ee
As we explained, the term with $j=0$ in this sum reproduces the operator $\hat l_0$,
and we finally come to the equation (we put $\mu=1$ here)
\be\label{eqnr2}
\left(-x{d\over dx} + \sum_{i=1}^{r-1} (r+1)i\,x^{(r+1)i}\cdot\hat O_{(r+1)i}\right) Z_r\{p\}=0\nn\\
\left[\hat l_0, \hat O_{(r+1)i}\right] = (r+1)\, i\cdot  \hat O_{(r+1)i}
\ee
The solution to this equation is the iterated integral
\be
 Z_r\{p\} =
1 \ +\ \int^1\! \hat A(t)\frac{dt}{t} \
+\ \int^1 \! \hat A(t)\frac{dt}{t} \int^{t}\!  \hat A(t')\frac{dt'}{t'} \ +\
 \int^1\!  \hat A(t)\frac{dt}{t} \int^{t}\!  \hat A(t')\frac{dt'}{t'}\int^{t'}\!  \hat A(t'')\frac{dt''}{t''}
 \ +\  \ldots
\ee
with $\hat A(t) := \sum_{k=2}^r  kt^{k}\hat O_{(r+1)k}$, $t:=x^{r+1}$,
i.e. the series
\be
 Z_r\{p\} = \sum_{s=1}\sum_{i_1,\ldots,i_s=1}^{r-1} \frac{i_1\ldots i_s\cdot\hat O_{(r+1)i_1}\ldots \hat O_{(r+1)i_s}}{(i_1+\ldots+i_s) \ldots   (i_{s-1}+i_s) i_s}
\label{Osum}
\ee
where some $i_k$ can be the same.
The coefficient is  the repeated integral
\be
\int^1_0 t^{i_1}\, \frac{dt}{t}  \int^t {t'}^{\,i_{2}}\,
\frac{dt'}{t'}\int^{t'} t''^{\,i_3}\,\frac{dt''}{t''}
\ \ldots \
\ee
In the commuting case, the coefficients would  sum up just to
\be
\sum_{\{k_a\}}\prod_{a=1}^{r-1}\frac{1}{k_a!}\hat O_{(r+1)i_a}^{k_a}
= \exp\left(\sum_{a=1}^{r-1}\hat O_{(r+1)i_a}\right)
\ee
but, in the generalized Kontsevich model, $\hat O_{i}$'s do not commute.
Still (\ref{Osum}) is a very explicit and practical expression,
we give some examples of its application in Appendix B.

\subsection{To $W$-representation from matrix Ward identity for GKM}

An interesting option would be to start directly from the identity \cite{MMM,GN,GKM,Mikh}
\be
\left\{V'\left(\frac{\p}{\p L^{tr}}\right) - L\right\}{\cal Z}_V = 0
\label{GN}
\ee
for the matrix integral
\be
{\cal Z}_V = \int dX e^{\tr (V(X) - V'(M)X)} =
\frac{e^{\tr M V'(M) - V(M)}}{\det V''(M)}\cdot Z_V\{p_k\}
\ee
from which we extract the GKM partition function $Z_V$
depending only on negative powers of the matrix variable $M$,
$p_k = \tr M^{-k}$.
For the monomial potentials $V_r(X) = \frac{X^{r+1}}{r+1}$, this means that $M^r=L$,
and $Z_V=Z_r$ turns out to be independent of all $p_{rn}$, see \cite{GKM,versus} for details.
In this case,
\be
{\cal Z}_r = \frac{e^{\frac{r}{r+1}\tr M^{r+1}}}
{\det \left(\sum_{i=0}^{r-1} M^i \otimes  M^{r-1-i}\right)}\cdot Z_r\{p_k\}
\ee
and substitution into (\ref{GN}) gives a sum of terms with $r+1$ different gradings,
associated with $r$ derivatives of the exponential.
If we multiply the equation by $M$ and take a trace,
the gradings (powers of $M^{-1}$) will be $n(r+1)$ with $n=-1,0,\ldots,(r-1)$.
Actually the lowest grading with $n=-1$ does not show up,
because
\be
M\cdot\left(\frac{e^{-\frac{r}{r+1}\tr M^{r+1}}\p\, e^{\frac{r}{r+1}\tr M^{r+1}}}
{\p (M^{tr})^r}\right)^r - M^{r+1}  = 0
\ee
The most interesting is grading $0$, where we get the operator $\hat l_0$.
Indeed,
\be
r\cdot  \tr \!\!\left\{M\cdot\left(\frac{e^{-\frac{r}{r+1}\tr M^{r+1}}\p\, e^{\frac{r}{r+1}\tr M^{r+1}}}
{\p (M^{tr})^r}\right)^{r-1} \frac{\p Z_r}{ \p (M^{tr})^r} \right\}
= \nn \\
= r\cdot \sum_k  \tr\!\! \left(M^r \frac{\p p_k}{\p  (M^{tr})^r}\right)\cdot \frac{\p Z_r}{\p p_k}
= \sum_k k\cdot \tr M^{-k}\cdot \frac{\p Z_r}{\p p_k} = \sum_k kp_k\frac{\p Z_r}{\p p_k} =
\hat l_0 Z_r
\ee
There are two other contributions in this grading,
which do not contain $p$-derivatives of $Z_r$:
one appears when the $L$ derivative acts on $\det V''(\Lambda)$ instead of $Z_r$,
another one, when two $L$ derivatives act twice on the same exponential.
Analysis in other gradings gets more involved and will be addressed elsewhere.

\section{Conclusion}

In this letter, we resolve a puzzle of the $W$-representation \cite{MSh}
for the monomial generalized Kontsevich models \cite{GKM} beyond the cubic case \cite{Alexcubic}.
As usual, the deviation from the standard situation appeared very simple
but unexpected and implies far-going consequences.
It turned out that the $W$-representation is not an ordinary exponential but
an {\it ordered} $P$-exponential
of a linear combination of {\it non-commuting} $W$-like operators of
{\it different} gradings.
We remind that, like many other  matrix models \cite{UFN3},
the GKM partition function is a KP $\tau$-function \cite{GKM},
thus what we observe is a striking appearance of $P$-exponential
in the field of {\it integrable} systems.
This brings the seemingly simple matrix models into a direct contact
with Yang-Mills theories, where the $P$-exponentials play the central role:
as predicted long ago, the non-Abelian nature has no conflict with integrability.

In the narrower field of matrix models {\it per se}, the $W$-representations
provide a truly effective method for generating as many terms of the
GKM partition function as one needs.
This opens new possibilities for study of these very interesting and
archetypical models.
Some details are still lacking, and we have not yet derived a truly closed expression
for arbitrary $r$, this is one of the simplest subjects for the future work.

\section*{Acknowledgements}

We are indebted to A. Alexandrov for pointing to us the paper \cite{Zhou},
which attempted to find a $W$-representation of the GKM partition functions.
Despite it overlooked non-commutativity of the relevant operators,
which led to an oversimplified anzatz for the $W$-representation,
that paper forced us to revisit the problem and overcome our prejudices.
Another origin of our paper is our recent activity on a systematic approach to
solving Virasoro-like constraints \cite{MMMR1,MMMR2,MMMR3},
and we are very grateful to R. Rashkov for collaboration.

This work was supported by the Russian Science Foundation (Grant No.21-12-00400).

\section*{Appendix A: General formula for $\hat W^{(k)}_n$ \label{ded}}

In this Appendix, we describe how one can obtain the relevant $W$-operators
$\phantom.{_r}\hat{W}_n{(k)}$ by the normal ordering of a
product of currents \cite{FKN2}. The main point is that the spectral curve for the monomial GKM model
$Z_r$ (which can be obtained from the corresponding loop equations \cite{AMMP}) is the $r$-sheeted covering of a sphere, which is clear both from the integrable hierarchy point of view
(since the system is described by the $r$-th reduction of the KP hierarchy) \cite{FKN2,GKM}, and from
the topological recursion point of view \cite{AMM}. This is why it is natural to define the current to be
\be
J(z):= \sum J_n z^{-n/r-1}
\ee
with the current modes given by (\ref{shr}).
This expression involves the $r$-th root of $z$, and, hence, one has to specify which of the roots is used (the sheet of the covering). We denote choosing the $m$-th root as $z_m$. One can arbitrarily choose the ordering of $z_m$, $m=1,\ldots,r$, but, for the sake of definiteness, we choose them to be $z_{m+1}^{1/r}=\exp\Big({2\pi i\over r}\Big)\cdot z_m^{1/r}$ and denote $z_1=z$.
Note that {\it integer} powers of $z$ are the same for all $z_m$, but, at the level of $J(z)$, the
arguments are all different, and one can use non-singular operator expansions.
At the same time, the final answer contains only integer powers of $z$.

Now the procedure of constructing the $\phantom.{_r}\hat{W}_n{(k)}$-operators consists of three steps.
\begin{itemize}
\item[{\bf 1.}] The starting point is an auxiliary operator
\be
\phantom.{_r}\hat W_{aux}^{(k)}(z)=\frac{(-1)^{k+1}}{r^k} \sum_{1\le m_1<m_2<\ldots<m_k\le r}J(z_{m_1})\ldots J(z_{m_k})
\ee
which is very simple and general, but not normally ordered.
\item[{\bf 2.}] One has to normally order $\phantom.{_r}\hat W_{aux}^{(k)}(z)$ in such a way that all positive current modes are moved to the right.
\item[{\bf 3.}] {\bf After normal ordering}, one has to omit all the current modes divisible by $r$:
$J_{nr}=0$ in order to finally obtain $\phantom.{_r}\hat{W}^{(k)}(z)$.

Now note that both $\phantom.{_r}\hat W_{aux}^{(k)}(z)$ and $\phantom.{_r}\hat W^{(k)}(z)$ are single-valued, and, hence, are expanded into integer powers of $z$. Thus, one generates $\phantom.{_r}\hat{W}_n^{(k)}$ as
\be
\phantom.{_r}\hat{W}^{(k)}(z)=\sum_n \phantom.{_r}\hat{W}_n^{(k)} z^{-n-k}
\ee
\end{itemize}

Now we demonstrate how this procedure works in a few examples.

\paragraph{\framebox{$\hat{W}_n^{(2)}$:}}

In this case, we have
\be\label{W2a}
\phantom.{_r}\hat W_{aux}^{(2)}(z)\ {\stackrel{step\ 1}{=}}\ -\frac{1}{r^2} \sum_{1\le m_1<m_2\le r}J(z_{m_1})J(z_{m_2})
=\nn\\
{\stackrel{step\ 2}{=}}\ -\frac{1}{r^2} \sum_{1\le m_1<m_2\le r}:J(z_{m_1})J(z_{m_2}):
-{1\over r^2}\sum_{1\le m_1<m_2\le r}\sum_{n_1,n_2>0}[J_{n_1},J_{-n_2}]z_{m_1}^{-n_1/r-1}z_{m_2}^{n_2/r-1}
=\nn\\
=-\frac{1}{r^2} \sum_{1\le m_1<m_2\le r}\ \sum_{n_1,n_2\in\mathbb{Z}}\omega_{m_1}^{-n_1}
\omega_{m_2}^{-n_2}:J_{n_1}J_{n_2}:z^{-(n_1+n_2)/r-2}+{r^2-1\over 24r}
\longrightarrow\nn\\
{\stackrel{step\ 3}{\longrightarrow}}\
\phantom.{_r}\hat W^{(2)}(z)=-\frac{1}{r^2} \sum_{1\le m_1<m_2\le r}\ \sum_{n_1,n_2\notin r\mathbb{Z}}\omega_{m_1}^{-n_1}
\omega_{m_2}^{-n_2}:J_{n_1}J_{n_2}:z^{-(n_1+n_2)/r-2}+{r^2-1\over 24r}
\ee
where we denoted $\omega_m:=\exp\Big({2\pi im\over r}\Big)$ and used that $[J_{n_1},J_{-n_2}]=n_1\delta_{n_1,n_2}$, and, hence, the anomaly term is
\be\label{an}
-{1\over r^2}\sum_{1\le m_1<m_2\le r}\ \sum_{n_1,n_2>0}[J_{n_1},J_{-n_2}]z_{m_1}^{-n_1/r-1}z_{m_2}^{-n_2/r-1}
=-{1\over r^2z^2}\sum_{1\le m_1<m_2\le r}\sum_{n>0} n\left({\omega_{m_2} \over\omega_{m_1}}\right)^{n}=\nn\\
=-{1\over r^2z^2}\sum_{1\le m_1<m_2\le r}{\omega_{m_1}\omega_{m_2}\over (\omega_{m_1}-\omega_{m_2})^2}
= {r^2-1\over 24r}{1\over z^2}
\ee
Calculating the sum over $n$ requires a regularization as usual for the anomaly.

At last, in order to rewrite the remaining sum in the last line of (\ref{W2a}), we use the identity
\be\label{i1}
{1\over 2}\left(\sum_{1\le m_1<m_2\le r}\omega_{m_1}^{-n_1}
\omega_{m_2}^{-n_2}+\sum_{1\le m_1<m_2\le r}\omega_{m_1}^{-n_2}
\omega_{m_2}^{-n_1}\right)=-{r\over 2}\delta_{n_1+n_2,rn}\ \ \ \ \ \ \hbox{for any }\ n_1,n_2\notin r\mathbb{Z}
\ee
so that we finally obtain
\be
\phantom.{_r}\hat W^{(2)}(z)={1\over 2r}\sum_{\substack{n_1,n_2\notin r\mathbb{Z}\\
n_1+n_2=rn}}{:J_{n_1}J_{n_2}:\over z^{n+2}}+
{r^2-1\over 24r}{1\over z^2}
\ee
which gives rise to (\ref{W2r}).

\paragraph{\framebox{$\hat{W}_n^{(3)}$:}}
In this case,
\be\label{W3a}
\phantom.{_r}\hat W_{aux}^{(3)}(z)\ {\stackrel{step\ 1}{=}}\ \frac{1}{r^3} \sum_{1\le m_1<m_2<m_3\le r}
J(z_{m_1})J(z_{m_2})J(z_{m_3})
=\nn\\
{\stackrel{step\ 2}{=}}\ \frac{1}{r^3} \sum_{1\le m_1<m_2<m_3\le r}:J(z_{m_1})J(z_{m_2})J(z_{m_3}):
+\hbox{terms linear in J(z)}
=\nn\\
=\frac{1}{r^3} \sum_{1\le m_1<m_2<m_3\le r}\ \sum_{n_1,n_2,n_3\in\mathbb{Z}}\omega_{m_1}^{-n_1}
\omega_{m_2}^{-n_2}\omega_{m_3}^{-n_3}:J_{n_1}J_{n_2}J_{n_3}:z^{-(n_1+n_2+n_3)/r-3}
+\hbox{terms linear in J(z)}
\longrightarrow\nn\\
{\stackrel{step\ 3}{\longrightarrow}}\
\phantom.{_r}\hat W^{(3)}(z)={1\over 3r^2}\sum_{\substack{n_1,n_2,n_3\notin r\mathbb{Z}\\
n_1+n_2+n_3=rn}}{:J_{n_1}J_{n_2}J_{n_3}:\over z^{n+3}}
\ee
The terms linear in $J$ in the second line are omitted at the third step, since they are proportional to $J_{nr}$. This is evident since the whole expression should be single-valued, and, hence, it depends only on integer powers of $z$, i.e. on $J_{nr}$. It can be manifestly seen in the following way: it is a sum of three terms of the form
\be
{1\over r^3}\sum_{1\le m_1<m_2<m_3\le r}\ \sum_{n_1,n_2>0,n_3}[J_{n_1},J_{-n_2}]J_{n_3}
z_{m_1}^{-n_1/r-1}z_{m_2}^{-n_2/r-1}z_{m_3}^{-n_3/r-1}=\nn\\
={1\over r^3z^3}\sum_{1\le m_1<m_2<m_3\le r}\sum_{n>0,n_3} n\left({\omega_{m_2} \over\omega_{m_1}}\right)^{n}
\omega_{m_3}^{-n_3}J_{n_3}z^{-n_3/r}={1\over r^3z^3}\sum_{n_3}z^{-n_3/r}J_{n_3}\sum_{1\le m_1<m_2<m_3\le r}{\omega_{m_1}\omega_{m_2}\omega_{m_3}^{-n_3}\over (\omega_{m_1}-\omega_{m_2})^2}\nn
\ee
with two other terms corresponding to $[J_{n_1},J_{-n_3}]J_{n_2}$ and $[J_{n_3},J_{-n_2}]J_{n_1}$. The sum of these three terms is proportional to
\be
\left(\sum_{1\le m_1<m_2<m_3\le r}+\sum_{1\le m_1<m_3<m_2\le r}+\sum_{1\le m_3<m_2<m_1\le r}\right)
{\omega_{m_1}\omega_{m_2}\over (\omega_{m_1}-\omega_{m_2})^2}\omega_{m_3}^{-n_3}=-{r(r^2-1)(r-2)\over 24}\delta_{n_3,rn}
\ee
giving rise to the sum
\be
-{(r^2-1)(r-2)\over 24r^2}\sum_nJ_{nr}z^{-n-3}
\ee

The last line in (\ref{W3a}) is due to the identity
\be\label{i2}
{1\over 6}\left(\sum_{1\le m_1<m_2<m_3\le r}\omega_{m_1}^{-n_1}
\omega_{m_2}^{-n_2}\omega_{m_3}^{-n_3}+\hbox{all permutations of }n_1,n_2,n_3\right)={r\over 3}\delta_{n_1+n_2+n_3,rn}\nn\\
\hbox{for any }\ n_1,n_2,n_3\notin r\mathbb{Z}
\ee

\paragraph{\framebox{$\hat{W}_n^{(4)}$:}}
In this case, the calculation is very similar, but there is a subtlety. That is, one needs a counterpart of formulas (\ref{i1}) and (\ref{i2}). Now it, however, has a more subtle structure: the r.h.s. depends not only on $r$ and on divisibility of the sum $n_1+n_2+n_3+n_4$ by $r$, but also on divisibility of pairs $n_1+n_2$, etc. Indeed, the identity is
\be\label{i4}
{1\over 24}\left(\sum_{1\le m_1<m_2<m_3<m_4\le r}\omega_{m_1}^{-n_1}
\omega_{m_2}^{-n_2}\omega_{m_3}^{-n_3}\omega_{m_4}^{-n_4}+\hbox{all permutations of }n_1,n_2,n_3,n_4\right)=\nn\\
=-{r\over 4}\Big(1-{c(n_1,n_2,n_3,n_4)r\over 6}\Big)\delta_{n_1+n_2+n_3+n_4,rn}\ \ \ \ \ \ \
\hbox{for any }\ n_1,n_2,n_3,n_4\notin r\mathbb{Z}
\ee
where the coefficient $c(n_1,n_2,n_3,n_4)$ is the number of different combinations of $n_i$'s with pairwise sums divisible by $r$. For instance, $c(1,1,3,3)=0$ (no combinations), $c(3,4,4,5)=1$ (1 combination), $c(3,3,5,5)=2$ and $c(2,2,2,2)=3$ at $r=8$. This immediately implies that the normally ordered quartic combination of currents turns into the difference of two terms:
\be
\phantom.{_r}\hat W^{(4)}(z)={1\over 4r^3}\sum_n {1\over z^{n+4}}
\left(\sum\limits_{n_1+n_2+n_3+n_4 = rn}  : J_{n_1} J_{n_2}J_{n_3}J_{n_4} :
- \dfrac{r}{2}  \sum\limits_{ \substack{
p+q=n
\\n_1+n_2 =rp
\\n_3+n_4=rq} } : J_{n_1} J_{n_2} J_{n_3} J_{n_4}  :\right) +\ldots
\ee
In order to calculate the terms of the form $:JJ:$, one proceeds similarly to (\ref{an}) and uses the identity
\be
\sum_{1\le m_1<m_2<m_3<m_4\le r} \hbox{Sym}_{\omega_{m_i}}
{\omega_{m_1}\omega_{m_2}\omega_{m_3}^{-n_1}\omega_{m_4}^{-n_2}\over (\omega_{m_1}-\omega_{m_2})^2}
= \left[r^2\cdot {r^2-1\over 12}-{r\over 2}\cdot\Big((n_1)_r^2+(n_2)_r^2-1\Big)\right]\delta_{n_1+n_2,rn}\nn\\
\hbox{for any }\ n_1,n_2\notin r\mathbb{Z}
\ee
Here the symmetrization symbol Sym$_{\omega_{m_i}}$ means that we sum over all permutations of $\omega_i$.

At last, in order to evaluate the remaining constant anomaly term, one twice uses the summation as in (\ref{an}), and the identity
\be
\sum_{1\le m_1<m_2<m_3<m_4\le r} \hbox{Sym}_{\omega_{m_i}}
{\omega_{m_1}\omega_{m_2}\over (\omega_{m_1}-\omega_{m_2})^2}
{\omega_{m_3}\omega_{m_4}\over (\omega_{m_3}-\omega_{m_4})^2}=8r\cdot{(r^2-1)(r-2)(r-3)(5r+7)\over 5760}
\ee
in order to ultimately obtain (\ref{W4}).

\paragraph{\framebox{$\hat{W}_n^{(5)}$:}}
Similarly, for spin 5 generators, one obtains
\be\label{i5}
{1\over 120}\left(\sum_{1\le m_1<m_2<m_3<m_4<m_5\le r}\omega_{m_1}^{-n_1}
\omega_{m_2}^{-n_2}\omega_{m_3}^{-n_3}\omega_{m_4}^{-n_4}\omega_{m_5}^{-n_5}+\hbox{all permutations of }n_1,n_2,n_3,n_4,n_5\right)=\nn\\
={r\over 5}\Big(1-{c(n_1,n_2,n_3,n_4,n_5)r\over 12}\Big)\delta_{n_1+n_2+n_3+n_4+n_5,rn}\ \ \ \ \ \ \
\hbox{for any }\ n_1,n_2,n_3,n_4,n_5\notin r\mathbb{Z}
\ee
and
\be
\phantom.{_r}\hat W^{(5)}(z)={1\over 5r^4}\sum_n {1\over z^{n+5}}
\left(\sum\limits_{n_1+n_2+n_3+n_4+n_5 = rn}  : J_{n_1} J_{n_2}J_{n_3}J_{n_4}J_{n_5} :
- \dfrac{5r}{6}  \sum\limits_{ \substack{
p+q=n
\\n_1+n_2+n_3 =rp
\\n_4+n_5=rq} } : J_{n_1} J_{n_2} J_{n_3} J_{n_4} J_{n_5} :\right) +\ldots
\ee
The term cubic in currents is obtained from the calculation similar to (\ref{an}) with help of the identity
\be
\sum_{1\le m_1<m_2<m_3<m_4<m_5\le r} \hbox{Sym}_{\omega_{m_i}}
{\omega_{m_1}\omega_{m_2}\omega_{m_3}^{-n_1}\omega_{m_4}^{-n_2}\omega_{m_5}^{-n_3}\over (\omega_{m_1}-\omega_{m_2})^2}&=&\nn\\
= \left(-r^2\cdot {r^2-1\over 6}+2r\cdot\left[(n_1)_r^2+(n_2)_r^2+(n_3)_r^2+r\Big(r-(n_1)_r-(n_2)_r-(n_3)_r\Big)-1\right]\right)&\delta_{n_1+n_2+n_3,rn}&\nn\\
 \hbox{for any }\ n_1,n_2,n_3\notin r\mathbb{Z}&&
\ee
This finally gives (\ref{W5}).

One can see that the way to evaluate the $W$-generators performed in this subsection is straightforward, but it makes computer calculations rather involved. However, all what one needs is knowledge of the sums
\be
\sum_{1\le m_1<\ldots<m_{2s+p}\le r} \hbox{Sym}_{\omega_{m_i}}\prod_{a=1}^s{\omega_{m_{2a-1}}\omega_{m_{2a}}\over (\omega_{m_{2a-1}}-\omega_{m_{2a}})^2}\prod_{b=1}^p\omega_{m_b}^{-n_b}
\ee

\section*{Appendix B:  Examples of GKM partition functions}

In this Appendix, we present the first orders of expansion of the partition functions $Z_2\{p\}$, $Z_3\{p\}$ and $Z_4\{p\}$ produced by the method described in the paper.
They are often needed in applications.
In this way, it is easy to generate many more terms: the number here is limited by the length
acceptable in a printed version.

{\footnotesize
\be
{\rm Cubic\ Kontsevich\ model:} \ \ \ \ \ \ \ \ \ \ \ \ \ \ \ \
Z_2\{p\}=1+x^3\left({p_3\over 48}+{p_1^3\over 12}\right)+x^6\left({25p_1^3p_3\over 576}+{25p_3^2\over 4608} +{p_1p_5\over 32}+ {p_1^6\over 288}\right)+\nn\\
+x^9\left({1225\over 55296}p_1^3p_3^2 +{49\over 13824}p_1^6p_3 +{7\over 384}p_1^4p_5 +{5\over 128}p_1^2p_7+{49\over 1536}p_1p_3p_5 +{35\over 3072}p_9+{p_1^9\over 10368}+{1225p_3^3\over 663552}\right)+\nn\\
+
x^{12}\left({89425\over 7962624} p_3^3 p_1^3+{3577\over 1327104} p_3^2 p_1^6+{73\over 497664} p_3 p_1^9+{13\over 9216} p_1^7 p_5+{17\over 1536} p_1^5 p_7+{49\over 2048} p_5^2 p_1^2+{29\over 2048} p_7 p_5+{1715\over 36864} p_9 p_1^3+\right.\nn\\
\left.+{2555\over 147456} p_9 p_3+{105\over 2048} p_{11} p_1+{365\over 6144} p_7 p_3 p_1^2+{p_1^{12}\over 497664}+{89425\over 127401984} p_3^4+{3577\over 147456} p_3^2 p_1 p_5+{511\over 18432} p_3 p_1^4 p_5\right)
+\nn\\ \!\!\!\!\!\!\!\!\!\!
+x^{15}\left({7427\over 884736} p_9 p_1^6+{1169\over 122880} p_1^5 p_5^2+{19p_5 p_1^{10}\over 331776} +{247835\over 14155776} p_9 p_3^2+{8674225\over 1528823808} p_3^4 p_1^3+{346969\over 191102976} p_3^3 p_1^6+{7081\over 47775744} p_3^2 p_1^9+\right.\nn\\
\left.+{455\over 8192} p_{11} p_1^4+{29\over 36864} p_7 p_1^8
+{97p_1^{12} p_3\over 23887872} +{3395\over 32768} p_{11} p_3 p_1+{7427\over 98304} p_9 p_5 p_1+{35405\over 589824} p_7 p_3^2 p_1^2+{1649\over 73728} p_7 p_3 p_1^5+\right.\nn\\
\left.+{2813\over 98304} p_7 p_3 p_5+{166355\over 1769472} p_9 p_1^3 p_3+{49567\over 1769472} p_5 p_3^2 p_1^4 +{21\over 5120} p_5^3
+{1261\over 442368} p_5 p_3 p_1^7
+{4753\over 98304} p_3 p_1^2 p_5^2+{1739\over 24576} p_5 p_1^3 p_7+\right.\nn\\
\left.+{346969\over 21233664} p_5 p_3^3 p_1+{1155\over 8192} p_{13} p_1^2+{145\over 4096} p_7^2 p_1+{5005\over 98304} p_{15}
+{1734845\over 6115295232} p_3^5+{p_1^{15}\over 29859840}\right) +\ O(x^{18})\nn
\ee
}
{\footnotesize
\be
Z_3\{p\}=1+x^4\left(\frac 1{36} p_4+\frac 16 p_1^2p_2\right)
+x^8\left({13\over 216} p_1^2 p_2 p_4+{13\over 2592} p_4^2-{1\over 216} p_2^4+{1\over 72} p_1^4 {p_2^2}+{1\over 27} p_1^3 p_5+{1\over 27} p_1 p_7\right)+\nn\\
+x^{12}\left({325\over 279936} p_4^3-{5\over 324} p_2^2 p_8-{1\over 81} p_1 {p_2^3} p_5+{1\over 162} p_1^5 p_5 p_2-{1\over 1296} p_1^2 {p_2^5}+{25\over 972} p_4 {p_1^3} p_5-{1\over 162} p_2 {p_5^2}+{25\over 972} p_4 p_1 p_7-\right.\nn\\ -\left.{25\over 7776} p_2^4 p_4+{5\over 324} p_8 {p_1^4}+{7\over 162} p_1^2 p_{10}+{5\over 162} p_1^3 p_2 p_7+{325\over 15552} p_1^2 p_2 {p_4^2}+{25\over 2592} p_1^4 {p_2^2} p_4+{1\over 1296} p_1^6 {p_2^3}\right)+\nn\\
+x^{16}\left({12025\over 40310784} p_4^4+{25\over 1458} p_1^2 {p_7^2}+{35\over 729} p_1^3 p_{13}-{925\over 559872} p_2^4 {p_4^2}-{55\over 1458} p_2 p_{14}+{7\over 324} p_1^4 p_2 p_{10}+{925\over 69984} p_4^2 p_1 p_7-\right.\nn\\ -\left.{1\over 81} p_2^2 p_5 p_7-{185\over 11664} p_2^2 p_4 p_8+{925\over 69984} p_1^3 {p_4^2} p_5-{13\over 972} p_1^2 {p_2^2} {p_5^2}-{25\over 5832} p_1 {p_2^4} p_7-{25\over 1944} p_1^2 {p_2^3} p_8+{259\over 5832} p_4 {p_1^2} p_{10}-\right.\nn\\ -\left.{10\over 243} p_1 {p_2^2} p_{11}+{1\over 216} p_1^5 {p_2^2} p_7+{12025\over 1679616} p_1^2 p_2 {p_4^3}+{37\over 5832} p_1^5 p_2 p_4 p_5-{37\over 2916} p_1 {p_2^3} p_4 p_5+{185\over 5832} p_1^3 p_2 p_4 p_7-\right.\nn\\ -\left.{10\over 243} p_1 p_2 p_5 p_8-{1\over 243} p_1 {p_5^3}+{2\over 243} p_{11} {p_1^5}+{5\over 1944} p_8 {p_1^6} p_2+{10\over 729} p_1^4 p_5 p_7-{13\over 5832} p_1^3 p_5 {p_2^4}+{1\over 1944} p_1^7 p_5 {p_2^2}+\right.\nn\\ +\left.{185\over 11664} p_4 p_8 {p_1^4}+{1\over 1458} p_1^6 {p_5^2}+{1\over 31104} p_1^8 {p_2^4}-{7\over 729} p_2^3 p_{10}-{11\over 729} p_{11} p_5-{1\over 15552} p_1^4 {p_2^6}-{37\over 5832} p_2 p_4 {p_5^2}+\right.\nn\\ +\left.{37\over 46656} p_1^6 {p_2^3} p_4-{37\over 46656} p_1^2 {p_2^5} p_4+{925\over 186624} p_1^4 {p_2^2} {p_4^2}+{1\over 93312} p_2^8-{85\over 11664} p_8^2\right)+O(x^{20})\nn
\ee

\begin{equation}
\begin{split}
Z_{4}\{p\}=&1+x^5 \left(\frac{p_5}{32}+\frac{1}{8}p_3 p_1^2+\frac{1}{8} p_2^2p_1\right) + \\
   +& x^{10}
   \left(\frac{1}{128} p_3^2
   p_1^4+\frac{1}{64} p_2^2 p_3
   p_1^3+\frac{1}{32} p_7
   p_1^3+\frac{1}{128} p_2^4
   p_1^2+\frac{9}{256} p_3 p_5
   p_1^2+\frac{1}{16} p_2 p_6
   p_1^2+\frac{9}{256} p_2^2 p_5 p_1+\frac{5
   p_9 p_1}{128}-\frac{1}{64} p_2^2
   p_3^2+\frac{9 p_5^2}{2048}-\frac{p_3
   p_7}{128}\right) +
   \\
   &+ x^{15} \left(\frac{p_3^3
   p_1^6}{3072}+\frac{p_2^2 p_3^2
   p_1^5}{1024}+\frac{1}{256} p_3 p_7
   p_1^5+\frac{p_2^4 p_3
   p_1^4}{1024}+\frac{17 p_3^2 p_5
   p_1^4}{4096}+\frac{1}{128} p_2 p_3 p_6
   p_1^4+\frac{1}{256} p_2^2 p_7
   p_1^4+\frac{7}{512} p_{11}
   p_1^4+\frac{p_2^6
   p_1^3}{3072}+\frac{1}{96} p_6^2
   p_1^3+
\right.  \\ & \left. \phantom{x^{15}}+
   \frac{17 p_2^2 p_3 p_5
   p_1^3}{2048}+\frac{1}{128} p_2^3 p_6
   p_1^3+\frac{17 p_5 p_7
   p_1^3}{1024}+\frac{55 p_3 p_9
   p_1^3}{3072}+\frac{1}{32} p_2 p_{10}
   p_1^3-\frac{1}{512} p_2^2 p_3^3
   p_1^2+\frac{153 p_3 p_5^2
   p_1^2}{16384}+\frac{17 p_2^4 p_5
   p_1^2}{4096}+\frac{17}{512} p_2 p_5 p_6
   p_1^2-
\right.  \\ & \left. \phantom{x^{15}}-
   \frac{p_3^2 p_7
   p_1^2}{1024}+\frac{25 p_2^2 p_9
   p_1^2}{1024}+\frac{45 p_{13}
   p_1^2}{1024}-\frac{1}{512} p_2^4 p_3^2
   p_1+\frac{153 p_2^2 p_5^2
   p_1}{16384}-\frac{3}{512} p_7^2
   p_1-\frac{1}{64} p_2 p_3^2 p_6
   p_1-\frac{25 p_2^2 p_3 p_7
   p_1}{1024}+\frac{85 p_5 p_9
   p_1}{4096}-
\right.  \\ & \left. \phantom{x^{15}}-
   \frac{7}{512} p_3 p_{11}
   p_1+\frac{p_3^5}{3840}+\frac{51
   p_5^3}{65536}-\frac{19 p_3
   p_6^2}{3840}-\frac{17 p_2^2 p_3^2
   p_5}{2048}-\frac{1}{96} p_2^3 p_3
   p_6-\frac{1}{512} p_2^4 p_7-\frac{17 p_3
   p_5 p_7}{4096}-\frac{29 p_2 p_6
   p_7}{1920}-\frac{5 p_3^2
   p_9}{1536}-
\right.  \\ & \left. \phantom{x^{15}}-
   \frac{59 p_2 p_3
   p_{10}}{1920}-\frac{311 p_2^2
   p_{11}}{15360}-\frac{693
   p_{15}}{40960}\right)
  \ + \
   O(x^{20})
\end{split}  \nn
\end{equation}
}
{\footnotesize
\be
Z_5\{p\}& =&
1+x^6\left(\frac{p_2^3}{30}+\frac{1}{5} p_1 p_3
   p_2+\frac{1}{10} p_1^2
   p_4+\frac{p_6}{30} \right)
\ + \nn\\
&+& x^{12}\left(
\frac{p_2^6}{1800}+\frac{1}{150} p_1 p_3
   p_2^4+\frac{1}{300} p_1^2 p_4
   p_2^3+\frac{7}{900} p_6 p_2^3+\frac{1}{50}
   p_1^2 p_3^2 p_2^2-\frac{1}{100} p_4^2
   p_2^2+\frac{1}{25} p_1 p_7
   p_2^2-\frac{1}{50} p_3^2 p_4
   p_2+
   \frac{1}{50} p_1^3 p_3 p_4
   p_2
 \right. +\nn \\   &+&
   \frac{7}{150} p_1 p_3 p_6
   p_2+
   \frac{3}{50} p_1^2 p_8
   p_2-\frac{p_3^4}{300}+\frac{1}{200} p_1^4
   p_4^2+\frac{7 p_6^2}{1800}+\frac{7}{300}
   p_1^2 p_4 p_6+\frac{1}{25} p_1^2 p_3
   p_7-\frac{p_4 p_8}{100}+\frac{2}{75} p_1^3
   p_9-\frac{p_3 p_9}{75}+\frac{p_1
   p_{11}}{25}
   \left.
 \right)+\nn\\
 &+& O(x^{18})
\ee
\be
Z_6\{p\} =
1+x^7\left( \frac{5}{144}p_7+ \frac{1}{12}p_1^2p_5+\frac{1}{6}p_1p_2p_4
+\frac{1}{12}p_1p_3^2+\frac{1}{12}p_2^2p_3\right)
\ + \ O(x^{14})
\nn \\ \nn \\
\ldots \ \ \ \  \ \ \ \ \ \ \ \ \ \ \ \ \ \ \ \ \
\ \ \ \  \ \ \ \ \ \ \ \ \ \ \ \ \ \ \ \ \ \ \ \  \ \ \ \ \ \ \ \ \ \ \ \ \ \ \ \ \nn
\ee
}

In general
{\footnotesize\be
Z_{r}\{p\} = 1 + x^{r+1}\left(\frac{r-1}{24 r}p_{r+1}+  \frac{3^{-\delta_{r,2}}}{2r} p_1^2 p_{r-1}
+ \frac{2^{-\delta_{r,4}}\cdot\theta_{r>3}}{r} p_1p_2p_{r-2}
\ +\ \ldots
\right)
\ + \ \ \ \ \   \nn \\
+   x^{2r+2}\left( \frac{r^2-1}{24r^2} p_1p_{2r+1} +
\frac{(r-1)(r+23)}{9\cdot 128\cdot r^2}p_{r+1}^2
+  \frac{(r-1)\theta_{r>2}}{6r^2} p_1^3p_{2r-1}
+  \frac{3^{-\delta_{r,2}}\cdot(r+23)}{12\cdot (2r)^2} p_1^2p_{r-1}p_{r+1}
+ \frac{3^{-2\delta_{r,2}}}{2\cdot (2r)^2} p_1^4p_{r-1}^2
+ \ldots
\right)
+ \nn \\
+  \ O\Big(x^{3r+3}\Big) \ \ \ \ \ \ \ \ \nn
\ee
}

\noindent
Omitted items depend on selection rules for $r$, i.e. enter with the Heaviside functions
like $\theta_{r>3}$ in the first bracket.


\begin{thebibliography}{12}

\bibitem{UFN3} A. Morozov,
Phys.Usp.(UFN) {\bf 37} (1994) 1;
hep-th/9502091; hep-th/0502010\\
A. Mironov, Int.J.Mod.Phys. {\bf A9} (1994) 4355; Phys.Part.Nucl.
{\bf 33} (2002) 537; hep-th/9409190

\bibitem{Vircon} F. David, Mod.Phys.Lett. {\bf A5} (1990) 1019\\
A. Mironov, A. Morozov, Phys.Lett. {\bf B252} (1990) 47-52\\
J. Ambj{\o}rn, Yu. Makeenko, Mod.Phys.Lett. {\bf A5} (1990) 1753\\
H. Itoyama, Y. Matsuo, Phys.Lett. {\bf 255B} (1991) 20

\bibitem{AMM} A.~Alexandrov, A.~Mironov, A.~Morozov,
  Int.\ J.\ Mod.\ Phys.\ A {\bf 19} (2004) 4127,
hep-th/0310113

\bibitem{Max1} L.~Cassia, R.~Lodin, M.~Zabzine,
  JHEP {\bf 2010} (2020) 126,
arXiv:2007.10354

\bibitem{MSh} A.~Morozov, S.~Shakirov,
  JHEP {\bf 0904} (2009) 064,
arXiv:0902.2627

\bibitem{wrep} A. Givental, 
math.AG/0008067

\bibitem{wrep1} A. Alexandrov, A. Mironov, A. Morozov,
Physica {\bf D235} (2007) 126-167, hep-th/0608228\\
A.~~Alexandrov, A.~Mironov, A.~Morozov,
Theor. Math. Phys. \textbf{150} (2007) 153-164,
hep-th/0605171

\bibitem{wrep2} A.Okounkov,
Math.Res.Lett. {\bf 7}
(2000) 447-453;\\
V.Bouchard, M.Marino,
In: {\sl From Hodge Theory to Integrability and tQFT: tt*-geometry},
Proceedings of Symposia in Pure Mathematics, AMS (2008), arXiv:0709.1458;\\
S.Lando,
In: {\sl Applications of Group Theory to Combinatorics}, Koolen,
Kwak and Xu, Eds.
Taylor \& Francis Group, London, 2008, 109-132;\\
M.Kazarian,
arXiv:0809.3263;\\
A.Mironov, A.Morozov,
JHEP \textbf{0902} (2009) 024, arXiv:0807.2843

\bibitem{MMMR2} A.~Mironov, V.~Mishnyakov, A.~Morozov, R.~Rashkov,
arXiv:2105.09920

\bibitem{Max2}  L.~Cassia, R.~Lodin, M.~Zabzine,
  arXiv:2102.05682

\bibitem{Alexcubic} A. Alexandrov, Mod.Phys.Lett. {\bf A26} (2011) 2193-2199, arXiv:1009.4887

\bibitem{otherW} A.~Alexandrov,
  Adv.Theor.Math.Phys.\  {\bf 22} (2018) 1347,
arXiv:1608.01627\\
H.~Itoyama, A.~Mironov, A.~Morozov,
  JHEP {\bf 1706} (2017) 115,
arXiv:1704.08648\\
A.~Mironov, A.~Morozov,
  Phys.\ Lett.\ B {\bf 771} (2017) 503,
arXiv:1705.00976

\bibitem{GKM} S.~Kharchev, A.~Marshakov, A.~Mironov, A.~Morozov, A.~Zabrodin,
  Phys.Lett. {\bf B275} (1992) 311,
  hep-th/9111037\\
S.~Kharchev, A.~Marshakov, A.~Mironov, A.~Morozov, A.~Zabrodin,
  Nucl.Phys.\ {\bf B380} (1992) 181,
  hep-th/9201013

\bibitem{versus} S. Kharchev, A. Marshakov, A. Mironov, A. Morozov,
Nucl.Phys. {\bf B397} (1993) 339-378, hep-th/9203043

\bibitem{GKMU} A.~Mironov, A.~Morozov, G.W.~Semenoff,
  Int.J.Mod.Phys. {\bf A11} (1996) 5031,
hep-th/9404005

\bibitem{Kon} M.~Kontsevich,
  Commun.Math.Phys.\  {\bf 147} (1992) 1

\bibitem{MMM} A.~Marshakov, A.~Mironov, A.~Morozov,
  Phys.Lett. {\bf B274} (1992) 280,

\bibitem{W} E.Witten, {\sl On the Kontsevich model and other models of
two-dimensional gravity}, in: New York 1991 Proc., Differential geometric
methods in theoretical physics, v.1, pp.176-216

\bibitem{AMMP} A.~Alexandrov, A.~Mironov, A.~Morozov, P.~Putrov,
  Int.J.Mod.Phys. {\bf A24} (2009) 4939,
arXiv:0811.2825

\bibitem{FKN1} M.~Fukuma, H.~Kawai, R.~Nakayama,
  Int.\ J.\ Mod.\ Phys.\ {\bf A6} (1991) 1385

\bibitem{Mikh} A.~Mikhailov,
  Int.\ J.\ Mod.\ Phys.\ {\bf A9} (1994) 873,
  hep-th/9303129

\bibitem{MP} A.~Mironov, S.~Pakulyak,
Theor. Math. Phys. \textbf{95} (1993) 604-625,
hep-th/9209100

\bibitem{KMMMP} S. Kharchev, A. Marshakov, A. Mironov, A. Morozov, S. Pakuliak,
Nucl.Phys. {\bf B404} (1993) 717-750, hep-th/9208044

\bibitem{Zhou} Jian Zhou,   
arXiv:1305.6991

\bibitem{FKN2} M.Fukuma, H.Kawai, R.Nakayama, 
Comm.Math.Phys. {\bf 143} (1992) 371-403

\bibitem{MMQgen} A.~Mironov, A.~Morozov,
  arXiv:2101.08759

\bibitem{GN} D. Gross, M. Newman,
Nucl.Phys. {\bf B380} (1992) 168-180

\bibitem{MMMR1} A.~Mironov, V.~Mishnyakov, A.~Morozov, R.~Rashkov,
JETP Letters {\bf 113:11} (2021),
arXiv:2104.11550

\bibitem{MMMR3} A.~Mironov, V.~Mishnyakov, A.~Morozov, R.~Rashkov, to appear
\end{thebibliography}
\end{document}